\documentstyle[prd,aps,psfig]{revtex}
\begin{document}     
\draft
\title{The Energy and Zenith Angle Distribution of 
Upward Going Muons and Neutrino Oscillations}
\author{Paolo Lipari and Maurizio Lusignoli}
\address{\it Dipartimento di Fisica, Universit\`a di Roma ``la Sapienza",\\
 and I.N.F.N., Sezione di Roma, Piazzale A. Moro 2,\\ I-00185 Roma, Italy}
\date{Received 5 December 1997}
\maketitle
\begin{abstract}
The energy and  zenith angle distribution of 
neutrino induced,  upward going  muons can give direct 
information on the presence of $\nu$--oscillations
in  precisely the range of parameters suggested as a solution of the 
atmospheric neutrino problem.  
We discuss here the uncertainties in the theoretical prediction.
The shape of the zenith angle  distribution of the muon  flux is 
quite insensitive to modifications 
of the theoretical input  and is a good  probe for the existence
of neutrino oscillations.
We conclude that the existing  data sample on $\nu$--induced  muons 
has the statistical power  to 
confirm or  refute the $\nu$--oscillation solution of the atmospheric
neutrino problem.

\end{abstract} 
\vspace{\baselineskip}
The recent results of the Superkamiokande (SK) experiment \cite {SK}
on atmospheric neutrinos
confirm the existence of a problem \cite{atm-problem}
in the relative numbers of $\nu_e$ and
$\nu_\mu$ induced events, that may be interpreted as an indication of
neutrino oscillations with large mixing. 
The favoured $\Delta m^2$
has however been shifted to values lower than those suggested by the
older Kamiokande results
\cite {Kam}.
The results of the Chooz experiment
\cite{Chooz} exclude 
large  amplitude oscillations 
of electron antineutrinos into other antineutrinos 
in the region of  interest for $\Delta m^2$, however 
$\nu_\mu \leftrightarrow \nu_\tau$ (or 
$\nu_\mu \leftrightarrow \nu_s$  where  $\nu_s$  is a sterile neutrino) 
oscillations remain a viable  solution of the problem.
A dedicated long--baseline experiment could provide 
a clear confirmation  of  the existence of oscillations \cite{foot_LBL};
the actual detection of $\tau$--leptons  produced in 
the detector would  be  especially striking,
however this would only be possible with a high energy neutrino beam and the
distance needed to observe such an effect 
would become prohibitively long if $\Delta m^2 \lesssim 10^{-3}\;\rm{eV}^2$.
It is therefore very  desirable to obtain clearer indications 
on the value of the parameter $\Delta m^2$. 

In this work, we rediscuss the information that can be obtained
from the analysis of $\nu$--induced upward going muons.
In particular we consider in detail the shape 
of the zenith angle distribution and suggest that it is 
a sensitive and model independent probe of the  existence 
of neutrino oscillations.
In the presence  of  $\nu$--oscillations the muon flux  is  reduced
in its  absolute value, its energy spectrum is distorted,
and  the zenith angle  distribution is deformed.
 Since the $\nu$ induced flux can be measured
only in half of the solid  angle, 
and the  deviations from the up--down  symmetry
$\phi_\mu (E,\theta) = \phi_\mu(E,\pi-\theta)$,
valid in the absence of oscillations,
cannot be measured, the interpretation of the data  requires a comparison 
with a calculated prediction, and the estimate of 
theoretical uncertainties in the calculation is crucial.

The absolute  value  of the predicted flux is  affected by a large 
uncertainty $\sim 20\%$ \cite{Frati93}.  
This  uncertainty  is not relevant in the  study 
of the shape of the muon  energy spectrum.
Since no underground muon  detector has used a magnetic  field,
a measurement of the energy
can only be obtained  for the muons that stop in the detector
using the particle range. 
In the presence of oscillations  the  flux  of low energy
muons is  suppressed more, and the  ratio $r = N_s/N_p$  of  the numbers
of  `stopping' and `passing' muon events  is reduced
with respect to the no--oscillation prediction.
From a  measurement of $r$, the IMB collaboration
\cite{IMB-upmu} has  obtained  an  {\em exclusion} region  in the 
$(\Delta m^2, \sin^2 2\,\theta_{\mu\tau})$ plane 
that is in sharp  contradiction with the 
{\em allowed} region obtained  by SK \cite {SK} from the analysis  of 
the events with $\nu$ interactions inside the detector.
The significance of the discrepancy  is much reduced  in the critical
reanalysis  of \cite{LLS}, where  the prediction for
the  value of $r$ in the absence of oscillations
is increased by  $\sim 12\%$  with respect 
to the calculation  used in the original IMB work.
The SK detector, with a larger effective area, larger  containement
volume  and  better spatial and energy resolution  can 
provide a new and more accurate measurement of the
stop/pass  ratio.

The direction of the muons 
is  well  measured  in  underground  detectors
and the shape of the angular  distribution can be studied in detail
and  compared  to   different predictions (with or  without  oscillations)
using  statistical  tests.
For illustration and the sake of clarity
in this  work we will  simply use 
the ratio of the of `horizontal' and  `vertical' muon events,
a single  number, 
to describe the shape of the distribution:
\begin{equation}
S = 
\int_{- {1\over 2} }^{0} d\cos \theta 
\,{dN \over d\cos \theta} 
\cdot
\left [\int_{-1}^{-{1\over 2}} d\cos\theta 
\, {dN \over d\cos \theta} \right]^{-1}.
\end{equation}
In the presence of oscillations the flux of vertical muons is
supressed  more (because the  quasi--collinear parent  neutrinos
have longer pathlengths) and $S$ increases,
by as much as 30\% for 
$\Delta m^2$ in the  region of interest and maximal  mixing.
The separate  measurement 
of the quantities $S_s$ and $S_p$, related to 
the angular distributions
for `stopping' and `passing' muons, 
results in  an improvement of  the sensitivity,
a better control of systematic effects, and  in case of   a positive  signal 
a more precise indication of the value of $\Delta m^2$.
We estimate the  uncertainty in the predicted 
value of $S$, $S_{s}$ and $S_p$  as  $\lesssim 4\%$.

In previous theoretical   discussions of 
stopping muons \cite{ALL93,Frati93}
it was assumed  for simplicity that 
stopping  muons are in the energy interval 
$E_{\min} \le E_\mu \le E^*$, and  passing  muons have  
energies $E_\mu \ge E^*$.
This  approximation is clearly not  realistic.
For simple geometrical reasons, the energy interval  where
stopping  muons can be identified depends on the zenith angle;
moreover in general, even for 
a fixed zenith angle, muons of the same energy can either stop in 
or pass through the detector, depending on their trajectory.
In general the  rates of stopping and passing  muons in  the detector
can be estimated as:
\begin{equation}
{dN_{s(p)} \over d\cos\theta} =
\int_0^\infty dE_\mu ~ \phi_\mu(E_\mu, \cos\theta) ~
A_{s(p)} (E_\mu, \theta) 
\label {eq:rate}
\end{equation}
where $\phi_\mu (E_\mu, \cos\theta)$ is the  differential 
flux of upward going muons
of energy $E_\mu$  and zenith angle $\theta$,
and $A_{s(p)} (E_\mu,\theta)$  is  the  detector acceptance
for stopping (passing)  muons.
A   detailed calculation of the  acceptances 
as a function of  muon energy  and  direction
is  only possible for the experimental   collaboration.
In the following, considering the case of
the SK  detector,
as a  first approximation we describe it
as a homogeneous cylindrical  volume of radius $R = 16.9$~m and height 
$H = 36.2$~m  \cite{SK-upmu} and  approximate the muon  trajectories   as
straight  lines  with length equal to the muon range in water;
in this  way the  calculation of the acceptance 
is  reduced  to  an elementary geometry problem.

The requirement that the muon track be entirely contained in the detector
volume gives:
\begin{equation}
A_{s} (E_\mu, \theta) = A(L(E_\mu), \theta) ,
\end{equation}
where  $L(E)$ is the muon range in water and
\begin{equation}
A(L, \theta) =
  2\,R\,H\,\sin\theta ~\sqrt {1 - x^2}
   +  2\,R^2 |\cos \theta| ~[ \cos^{-1} x 
   - 3\,x \,\sqrt {1 - x^2} ] \; \Theta \bigl[L_{max}(\theta) - L \bigr]\;,
\end{equation}
\noindent
with $x = L\sin\theta/2R$ and
$L_{\max}(\theta) = {\min }\,[2\,R/\sin\theta$, $H/|\cos\theta|]$.
The acceptance for stopping  muons  is non--zero only when 
the  muon range  is  in the interval
$L_{\min} \le L(E) \le L^*$,
where $L_{\min}$ is the 
minimum track length   required for 
detection   (7~m in the  first   analysis of SK \cite{SK-upmu}), and
$L^*$ is the maximum linear dimension
of the detector (49.5~m in SK); this  corresponds to  the
energy interval $1.45\le E_\mu \le 10.63$~GeV.
The acceptance for passing muons is given by:
\begin{equation}
A_{p} (E_\mu, \theta) = A(L_{\min}, \theta) - A_{s} (E_\mu, \theta) \;.
\end{equation}

In Table I we give our estimates of some quantities of interest
(total event rate, $r$, $S_s$, $S_p$) 
calculated  in the absence of oscillations  with different choices 
of the theoretical input,
as described in the first  three  columns.
Column~1 indicates the choice of  a neutrino flux
from those available  in the literature
\cite{Bartol,HKKM,Butk,Volkova,Lipari93,FLUKA,Mitsui}.
In column~2 the label LLS  indicates 
that the contributions of quasi--elastic  scattering
and of single pion production  to the neutrino
cross sections  have been  explicitely included \cite{LLS}, 
while the label DIS indicates  that the neutrino  cross section is 
described  by the formulae of deep inelastic  scattering;
the more accurate treatment of the
lowest multiplicity channels in the LLS option
results in a higher total flux and a higher stop/pass ratio $r$.
The  third  column indicates the choice made for the Parton
Distribution Functions (PDF), using the label defined in the
library code PDFLIB \cite{PDFLIB}.
The first row in table I 
contains the absolute values of the predictions for our reference
model (Bartol $\phi_\nu$  \cite{Bartol}, PDF by \cite{GRV}
and LLS \cite{LLS} cross sections);
in the other rows  we report  ratios
with respect to the reference values. 

Our  reference model  predicts  a rate of 521~yr$^{-1}$  of 
passing  muons with  track length larger  than 7~m. This can be compared 
with a preliminary result obtained  in  229  days of
live time of  $425 \pm 26$~yr$^{-1}$ \cite{SK-upmu}.
We also  predict a  rate of 218~yr$^{-1}$ stopping muons.
It is important  to observe that this prediction is quite sensitive
to the  value of the cut $L_{\min}$, because  the muon  flux
is steeply falling with energy in this  region;
a lower (higher) value $L_{\min} = 5$ (9)~m   results in a  
stopping rate of 269 (169)~yr$^{-1}$.
It will be therefore important to  consider in detail
the resolution in the measurement of the stopping point of the muon  tracks.
The distribution of  the length $L$ of the stopping  tracks is
a measurement of the muon spectrum  in the low  energy region;
however this  region is narrow, and with the expected event rates
the possible  distortions   of
the $dN/dL$ distributions  will be hard to  detect.

Our predictions for  the 
parameters $S_s$ and $S_p$ that describe the  shape of the zenith angle
distributions of stopping and passing muons are 1.45 and 1.63.
In both cases there is an excess of horizontal  muons,
and the anisotropy is stronger
for the higher energy (passing) muons, $S_p > S_s$.
The anisotropy   of the muon flux
and its  energy dependence 
originate from the zenith angle  dependence of the neutrino flux.
The angular  distribution is approximately isotropic
for low energy neutrinos, and it develops a stronger and stronger  dependence
on the zenith angle  with increasing  neutrino energy. This can be understood 
by considering 
the larger decay probability for
high energy mesons and muons traveling in a direction near to the horizontal. 

The spread of numbers in the 
columns  of table I can be used
as an estimate of the uncertainty  in the prediction
for the different observables.
We  estimate 
$\Delta r \sim 8\%$, $\Delta S_{s} \sim 3\;\%$ and 
$\Delta S_{p} \sim 4\;\%$.
The MMK \cite{Mitsui} neutrino flux  gives $S_{s(p)}$ 
larger than our reference value by a factor 1.16 (1.13), in contrast to
all other models, and we disregard this prediction in the following.

To explore the possible effects of 
modifications in the theoretical inputs of the neutrino flux  calculation,
we  have  repeated the $\nu$--flux calculation of 
\cite{Lipari93} including some extreme changes in the initial assumptions.
In \cite{Lipari93} the primary cosmic ray flux has a power law
energy spectrum $E^{-\alpha}$ with  $\alpha = 2.7$;
steepening (flattening)  the  spectrum 
using $\alpha = 2.8$ (2.6)  increases  (decreases) the 
stop/pass ratio $r$ by +26\% (--21\%),  correspondingly the $\cos \theta$
distribution  becomes flatter (steeper) with the much smaller  variation
of $\mp$1\% in $S_s$ and  $\mp$3\% in $S_p$.
Increasing (reducing) the K meson yield by a factor 1.5 changes 
$r$ by 
$\pm$8\%, $S_s$ by $\pm$3\%   and $S_p$ by $\pm$4\%.
These tests  indicate that the prediction on the shape of the 
angular distribution  is  quite stable. 

The effects of neutrino oscillations  on the observables that 
we are  discussing are illustrated in the figures.
In fig.~1 the solid curve
shows the dependence on $\Delta m^2$ for maximal mixing
of the ratio $r$ between the number of stopping and passing events,
for our reference model.
The error bars indicate {\sl only} statistical errors
estimated for two years of SK data--taking (i.e. in a few months from
now). 
In the  region  $10^{-3} \le  \Delta m^2 ({\rm eV^2}) \le 10^{-2}$,
favoured by the anomaly in the contained
and semi--contained  events, the stop/pass  ratio has a value
$r \lesssim 0.30$  to be compared with a no--oscillation prediction
0.42. It is apparent that in this  region of
$\Delta m^2$, for large mixing,  the 
statistical significance  of the effects of oscillations  is  clear
(as large as $\sim 7\;\sigma$), and the theoretical uncertainty of
the prediction is  likely to be the  most important source of  error.
With  the estimate of 8\% for the theoretical error
and reasonable systematic errors in the experiment, an effect should
be clearly visible if neutrino oscillations are indeed the reason of the
atmospheric neutrino anomaly. 
However, 
the suppression of $r$ is approximately the same for $\Delta m^2$ =
$10^{-3}$ or $10^{-2}$ eV$^2$ and therefore with an analysis of this quantity
it will not be possible to resolve the present uncertainty on $\Delta m^2$.
In fig.~1  we also  included (dashed line) 
the  $\Delta m^2$ dependence for maximal  mixing  of $r$ 
for $\nu_\mu$ oscillations  into sterile  neutrinos, the curve being shifted 
because  of the matter  effects \cite{ALL93}.

We have plotted in fig.~2 versus $\Delta m^2$,
again for maximal mixing, the quantities $S_s$ and $S_p$ that are
a measure of the shape of the angular distributions.
The error bars indicate the estimated statistical errors, which in this
case are likely to be the dominant source of
error.
In the presence of neutrino oscillations
both $S_s$ and $S_p$ are larger than  the no--oscillation
expectation because  of the larger suppression of the vertical flux. 
For a fixed  value of the mixing parameter, the  effect   of
oscillations on $S_s$ ($S_p$) has 
a broad maximum for $\Delta m^2 \simeq 0.7\cdot 10^{-3}$
($10^{-2}$)~eV$^2$.
Qualitatively we can  expect that the 
angular distortion  produced by $\nu$--oscillations  
is  maximum when 
\begin {equation}
\Delta m^2  \sim {2 \pi \over R_\oplus}~\langle E_\nu \rangle \;,
\end{equation}
where $R_\oplus$ is the earth radius and $\langle E_\nu\rangle$ is
the typical energy of the neutrinos that  produce the upward going muon
signal;
for smaller  values of $\Delta m^2$ the oscillations
do not  have time  to  develop,  for larger  values the effects of oscillations
suppress  the  flux  equally for all directions. 
The signals of passing and  stopping  muons are produced by 
neutrinos with median energy 
$\langle E_\nu^{s(p)} \rangle \simeq 8$(100)~GeV
and this is reflected in the  positions of the maxima in fig.~2.
We also note that for large mixing and 
$\Delta m^2 \sim 0.2$--$1\cdot 10^{-3}$~eV$^2$ one should have
$S_s > S_p$, an essentially model--independent signal of
oscillations, albeit with low statistical significance.
The matter effects, present in the $\nu_\mu
 \leftrightarrow \nu_s$ case, induce a detectable but different
deformation in the angular distribution; the parameters $S_{s,p}$ are not 
a good measure of the effect in this case, since the suppression for 
`horizontal' and `vertical' events are approximately equal, after 
integration over the solid angle. 

In fig.~3 we present in
the usual ($\Delta m^2, \;\sin^2 2\,\theta_{\mu\tau}$) plane
the regions that may be excluded at
95\% c.l. after two years of SK running
assuming  that  the  experiment  measures the central  value
of the expectations (without neutrino oscillations), 
by a measurement of $r$ (solid line),
$S_s$ (dashed) and $S_p$ (dot--dashed). In order to produce this plot, we 
(arbitrarily) combined quadratically the expected statistical error and 
the theoretical uncertainty.
Conversely, the effect of atmospheric $\nu_\mu \leftrightarrow \nu_\tau$ 
oscillations for representative parameter values, 
maximal mixing and
$\Delta m^2 =  10^{-3}$ (10$^{-2}$)~eV$^2$,  would result in
a reduction in the stop/pass ratio $r$ of  30 (26)\%, with a deviation of
3.2 (2.7) $\sigma$'s  including the theoretical  error,
and  shape parameters
$S_s = 2.14 \pm 0.27$ ($1.55 \pm 0.21$)  and
$S_p = 1.82 \pm  0.12$ ($2.13 \pm  0.17$), that
correspond to  2.5 and 1.6 (0.5 and 2.9) 
standard deviation effects.  
In both cases the signal should be detectable, and the analysis offers 
a rough measurement of the $\Delta m^2$  parameter.

Experimental results on upward going muons
have already been  presented.
Baksan \cite{Baksan-upmu},
MACRO \cite{MACRO-upmu}, Kamiokande \cite{Kam-upmu},
IMB \cite{IMB-upmu1,IMB-upmu}, and Superkamiokande  \cite{SK-upmu}, have  
collected large statistical  samples  of `passing' events. The results for 
the total rates are inconclusive, 
with respect to the oscillations, because of the large errors and theoretical
uncertainties.
All these experiments have presented 
measurements of the muon flux above a threshold energy as a function of 
the zenith angle.
We point out that the extraction of a detector 
independent flux from the measured passing muon rates
is both non  trivial and model dependent, because no detector has
a well defined and angle independent threshold, and one needs 
corrections that make use of the theoretical prediction. 
Since this is normally done assuming the absence of oscillations,
the fluxes can be used to test this hypothesis, but great care has to
be taken in the extraction of oscillation parameters. The comparison
of event rates with detailed MonteCarlo predictions is free from these
interpretation problems,
that is why we preferred to make a rough estimate of the acceptance and discuss
predictions of directly measurable quantities.
For similar reasons, it seems dangerous to reduce the data of different
experiments to a common threshold, as was done in a recent
analysis \cite{Fogli}.
The experimental results are summarized in table~II,
where we give the ratio $S_\phi^{exp}$ of the vertical 
and horizontal fluxes obtained from the published data, and our  
no--oscillation prediction $S_\phi^{th}$.
All experiments measure  
a value larger than the prediction. This could
be interpreted as an effect of 
neutrino oscillations, but such a  conclusion would be premature.
The fit of the shape of the angular distributions for 
the individual experiments
is poor even after the inclusion of oscillations; 
moreover a  comparison between  experiments
shows that the normalizations  and  detailed  shapes of the data are not
in good agreement with each other, suggesting the presence of 
unforeseen systematic effects (see \cite{Fogli} for a discussion).
Nonetheless the qualitative result 
of an excess of horizontal muons in all experiments 
is a hint that should  be studied in more detail.

In summary, a better measurement of upward going muons, as it
will be possible in the SuperKamiokande detector, can test 
the neutrino oscillation 
solution of the (semi)contained atmospheric neutrino problem.
Sensitive quantities are: the ratio of stopping to passing muons,
already noted and discussed, and the shapes of angular distributions,
that are affected by an even smaller theoretical uncertainty.

\newpage
\begin{table}
\caption {Predictions for observables related to upward going muon flux 
in the absence of oscillations for the SuperKamiokande detector.}
\begin {tabular} { l  c  l  c  c  c  c  }
$\phi_\nu$ & $\sigma_\nu$ & PDF & Rate (yr$^{-1})$ &  $r$ &
$S_s$   & $S_p$  \\
\hline
 Bartol       & LLS & GRV94LO          &     738 &   0.42 &   1.45 &   1.63  \\
\hline
 HKKM   & LLS & GRV94LO               &      0.94 &   0.98 &   1.00 &   1.00  \\
 BDZ    & LLS & GRV94LO               &      1.04 &   1.08 &   1.00 &   1.02  \\
 Volkova & LLS & GRV94LO               &      0.94 &   1.02 &   1.02 &   1.04  \\
 Lipari  & LLS & GRV94LO               &      0.95 &   0.98 &   1.01 &   1.00  \\
 FLUKA   & LLS & GRV94LO               &     1.00  &   0.93  &  1.01  & 1.02  \\
 MMK    & LLS & GRV94LO               &      1.00 &   1.09 &   1.16 &   1.13  \\
\hline
 Bartol & LLS & GRVLO      &      0.94 &   1.00 &   1.00 &   1.00  \\
 Bartol & LLS & GRV94DIS   &      0.98 &   0.99 &   1.00 &   1.00  \\
 Bartol & LLS & CTEQ4-23   &      1.05 &   0.98 &   1.00 &   1.00  \\
 Bartol & LLS & EHLQ2      &      0.90 &   1.02 &   1.00 &   1.00  \\
\hline
 Bartol & DIS & GRV94LO  &      0.99 &   0.96 &   1.00 &   1.00  \\
 Bartol & DIS & GRVLO    &      0.92 &   0.95 &   1.00 &   1.00  \\
 Bartol & DIS & EHLQ2    &      0.85 &   0.89 &   1.00 &   1.00  \\
\end{tabular}
\end{table}

\vspace {1.5 cm}

\narrowtext
\begin{table}
\caption {Measurements  of the shape of the zenith angle distribution
of the $\nu$--induced  muon flux. The theoretical expectation is  based
on our reference model.}
\begin {tabular} { l  c  c  c  }
Detector & $E_t$ (GeV) & $S_\phi^{exp}$ & $S_\phi^{th}$ \\
\hline
Baksan     & 1.0 &  $1.82 \pm  0.18 $  &1.50 \\
MACRO\tablenotemark[1]  & 1.0 &  $2.97 \pm  0.65 $  &1.50 \\
IMB        & 1.8 &  $1.79 \pm  0.18 $  &1.53 \\
Kamiokande & 3.0 &  $1.91 \pm  0.21 $  &1.56 \\ 
SK         & 6.0 &  $1.83 \pm  0.23 $  &1.60 \\
\end{tabular}
\tablenotetext[1] {\scriptsize
The Macro detector has a very low acceptance for near to horizontal
muons and the datum in the first angular bin has a very large error,
that propagates into the ratio of fluxes. If one neglects the first bin 
and extrapolates from the others
a ratio $S_\phi$ similar to the other rows is obtained.}
\end{table}
\widetext

\newpage
\begin{figure} [t]
\centerline{\psfig{figure=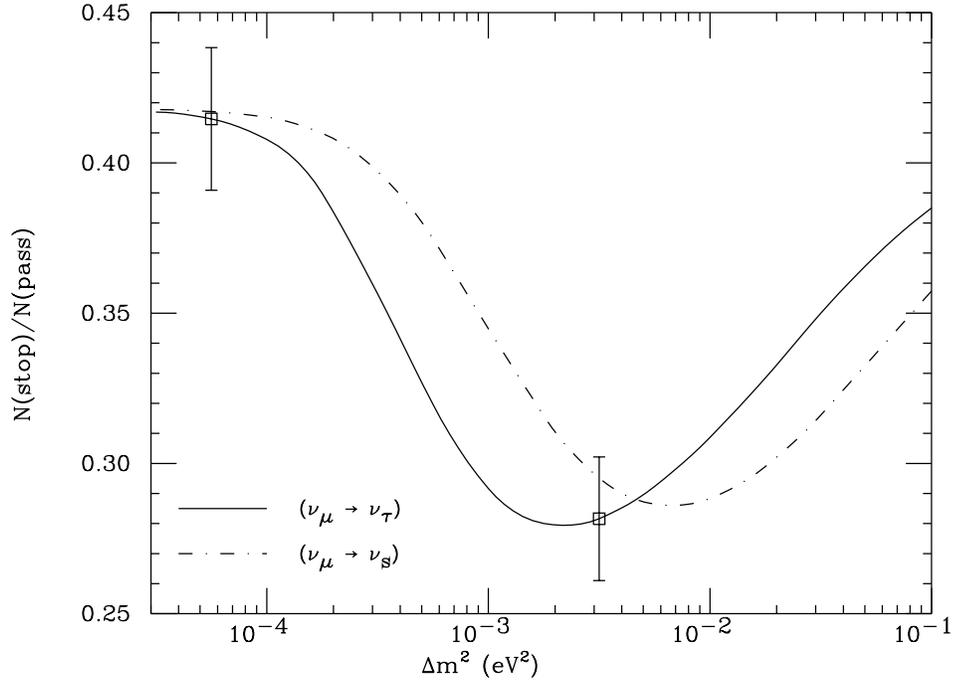,height=9cm}}
\vspace {0.4 cm}
\caption{Ratio of the  rates of `stopping' and `passing' muons
(with $L_{\min} = 7$~m)
in SuperKamiokande for  maximal  mixing.}
\end{figure}

\begin{figure} [t]
\centerline{\psfig{figure=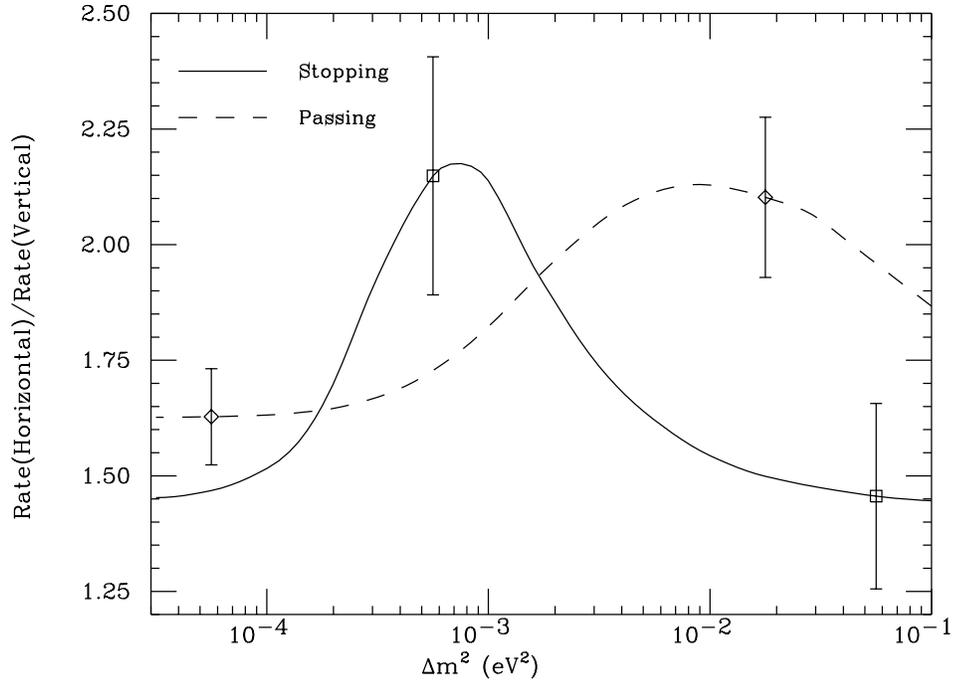,height=9cm}}
\vspace {0.4 cm}
\caption{Ratio of the  `horizontal' and `vertical'
muon  rates for stopping and passing events in
SuperKamiokande for $\nu_\mu$--$\nu_\tau$ oscillations with maximal mixing}
\end{figure}

\begin{figure} [t]
\centerline{\psfig{figure=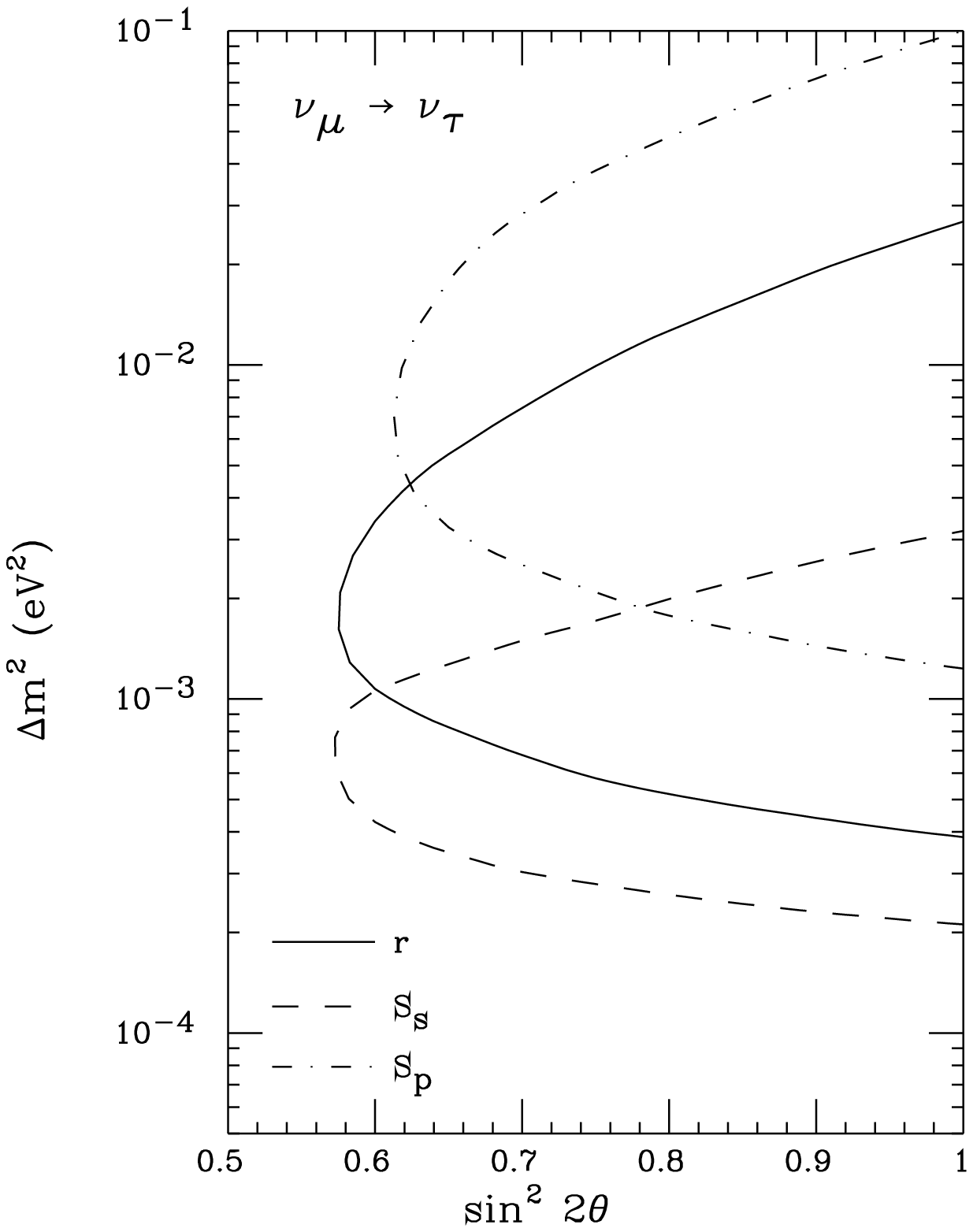,height=12cm}}
\vspace {0.4 cm}
\caption{Regions (to the right of the corresponding lines)
in the $(\Delta m^2, \sin^2 2\,\theta_{\mu\tau})$ plane,
that can be excluded at 95\% C.L.
with two years of SuperKamiokande  data
from measurements of $r$, $S_s$ ans $S_p$.
The  exclusion curves are calculated assuming
absence of oscillations and  including theoretical errors
of 8, 3 and 4\%  for the predictions.}
\end{figure}                                                


\begin{thebibliography}{999}
\bibitem {SK}  Y.\ Totsuka, to be published in the proceedings
of the Lepton Photon Conference, Hamburg, july 1997, 
available  at  http://www-sk.icrr.u-tokyo.ac.jp/doc/sk/pub/lp97.ps

\bibitem {atm-problem} See for example 
T.\ K.\ Gaisser, in
{\em Neutrino '96}, Proceedings of the 17th Conference
on Neutrino Physics and Astrophysics, Helsinki, 1996.


\bibitem {Kam}  Kamiokande  collaboration:
K.S. Hirata  {\it et al.}  Phys. Lett. B {\bf 205}, 416 (1988), 
{\it ibid.}  {\bf 280}, 146 (1992); 
Y. Fukuda {\it et al.} Phys. Lett. B {\bf 335}, 237 (1994).


\bibitem {Chooz} Chooz collaboration: M. Apollonio {\it et al.},  
 hep-ex/971102.

\bibitem {foot_LBL} 
For documentation on the numerous proposals, see: 
www.hep.anl.gov/NDK/Hypertext/long\_baseline.html

\bibitem{Frati93}
W.\ Frati, T.\ K.\ Gaisser, A.\ K.\ Mann, and T.\ Stanev,
Phys.\ Rev. D {\bf 48}, 1140 (1993).

\bibitem{IMB-upmu}
IMB Collaboration: R.\ Becker-Szendy {\em et al.},
Phys.\ Rev.\ Lett.\ {\bf 69}, 1010 (1992).

\bibitem {LLS}  P.\ Lipari, M.\ Lusignoli, and F.\ Sartogo,
		Phys.Rev.Lett. {\bf 74}, 4384 (1995).

\bibitem{ALL93} 
E.\ Akhmedov, P.\ Lipari, and M.\ Lusignoli,
Phys.\ Lett.\ B {\bf 300}, 128 (1993).  

\bibitem{SK-upmu}
SuperKamiokande Collaboration, 
presented by J.\ G.\ Learned in ICRC '97, Durban, South Africa,
astro-ph/9705197.

\bibitem{Bartol} 
[Bartol]
V.\ Agrawal, T.\ K.\ Gaisser, P.\ Lipari, and T.\ Stanev,
Phys.\ Rev.\ D {\bf 53} 1314, (1996).

\bibitem{HKKM}
[HKKM]
M.\ Honda, T.\ Kajita, K.\ Kasahara, and S.\ Midorikawa,
Phys.\ Rev.\ D {\bf 52}, 4985 (1995);

\bibitem{Butk}	
[BDZ]
A.\ V.\ Butkevich, L.\ G.\ Dedenko, and 
I.\ M.\ Zheleznykh,
Sov.\ J.\ Nucl.\ Phys.\ {\bf 50}, 90 (1989).

\bibitem{Volkova}
[Volkova]
L.V. Volkova,
Sov.J.Nucl.Phys.\ {\bf 31}, 784 (1980).

\bibitem{Lipari93}
[Lipari]
P.\ Lipari, Astropart.\ Phys.\ {\bf 1}, 195 (1993).

\bibitem{FLUKA}
[FLUKA]
G. Battistoni et al., to appear in the Proceedings of TAUP 97, LNGS,
and private communication. 

\bibitem{Mitsui}
[MMK]
K.\ Mitsui, Y.\ Minorikawa, and H.\ Komori,
Nuovo Cimento C {\bf 9}, 995 (1986).


\bibitem{Baksan-upmu}
Baksan Collaboration:
M.\ M.\ Boliev {\it et al.},
ICRC '95, Rome,
Vol~1, p.~686; {\em ibidem}, p.~722.

\bibitem{MACRO-upmu}
MACRO Collaboration: S.\ Ahlen {\em et al.},
Phys.\ Lett.\ B {\bf 357}, 481 (1995); 
F.\ Ronga {\em et al.}, in  
{\em Neutrino '96}, Proceedings of the 17th Conference
on Neutrino Physics and Astrophysics, Helsinki, 1996.

\bibitem{Kam-upmu}
Kamiokande Collaboration: A.\ Suzuki {\em et al.}
in Proceedings of 7th International Workshop on Neutrino 
Telescopes, Venice, 1996, edited by M.\ Baldo Ceolin
 p.~263.
 
\bibitem{IMB-upmu1}
IMB colllaboration:
D.\ W.\ Casper, in 
Proceedings of 3rd International Workshop on Neutrino 
Telescopes, Venice, 1991, edited by M.\ Baldo Ceolin
 p.~213.

\bibitem{PDFLIB}
H. Plotow-Besch, ``PDFLIB: The Parton Density Function 
Library,'' User's Manual, Version 7.07,
available at the URL http://consult.cern.ch/writeup/pdflib.

\bibitem{GRV}
M. Gluck, E. Reya, and A. Vogt,
Z.Phys. C {\bf 67}, 433 (1995).

\bibitem{Fogli}
G.L. Fogli, E. Lisi and A. Marrone,
preprint BARI-TH-280-97, hep-ph/9708213.

\end{thebibliography}
\end{document}